\documentclass{article}
\usepackage{amsfonts}
\usepackage{latexsym}
\usepackage{url}
\usepackage{theorem}
\usepackage[%
  breaklinks,
  colorlinks,
  bookmarks=false]
  {hyperref}

\newcommand{\N}{\mathbb{N}}
\newcommand{\eps}{\varepsilon}
\def\01{\{0, 1\}}

\newtheorem{theorem}{Theorem}
\newcommand{\thmref}[1]{\hyperref[#1]{Theorem~\ref*{#1}}}
\newtheorem{lemma}[theorem]{Lemma}
\newcommand{\lemref}[1]{\hyperref[#1]{Lemma~\ref*{#1}}}

\newenvironment{proof}[1][Proof.]{
	\par
	\noindent \textbf{#1}
}{
	\nobreak\leavevmode
	\hfill $\Box$\par\medskip
}
\newcommand{\eqref}[1]{\hyperref[#1]{(\ref*{#1})}}
\newcommand{\stepref}[1]{\hyperref[#1]{step~\ref*{#1}}}
\newcommand{\secref}[1]{\hyperref[#1]{Section~\ref*{#1}}}

\begin{document}

\title{Quantum Algorithms for Matching\\ and Network Flows}
\author{Andris Ambainis\thanks{%
Institute for Quantum Computing and Department of Combinatorics and
Optimization, University of Waterloo.  Supported by NSERC, CIAR and IQC University
Professorship.}\\
ambainis@math.uwaterloo.ca
\and
Robert {\v S}palek\thanks{%
CWI and University of Amsterdam.  Supported in part by the EU fifth framework
project RESQ, IST-2001-37559.  Work conducted while visiting University of
Waterloo and University of Calgary.}\\
sr@cwi.nl}
\date{}
\maketitle

\begin{abstract}
We present quantum algorithms for the following graph problems: finding a
maximal bipartite matching in time $O(n \sqrt{m+n} \log n)$, finding a maximal
non-bipartite matching in time $O(n^2 (\sqrt{m/n} + \log n) \log n)$, and
finding a maximal flow in an integer network in time $O(\min( n^{7/6} \sqrt m
\cdot U^{1/3}, \sqrt{n U} m) \log n)$, where $n$ is the number of vertices,
$m$ is the number of edges, and $U \le n^{1/4}$ is an upper bound on the
capacity of an edge.
\end{abstract}

\section{Introduction}

Network flows is one of the most studied problems in computer science.  We are
given a directed graph with two designated vertices: a source and a sink.
Each edge has assigned a capacity.  A network flow is an assignment of flows
to the edges such that the capacity of an edge is never exceeded and the total
incoming and outgoing flow are equal for each vertex except for the source and
the sink.  A size of the flow is the total flow going from the source.  The
task is to find a flow of maximal size.

After the pioneering work of Ford and Fulkerson \cite{ff:network-flows}, many
algorithms have been proposed.  Let $n$ denote the number of vertices and let
$m$ denote the number of edges.  For networks with real capacities, the
fastest algorithms run in time $O(n^3)$ \cite{karzanov:preflows,
3inds:maxflows}.  If the network is sparse, one can achieve a faster time $O(n
m (\log n)^2)$ \cite{gn:maxflows}.  If all capacities are integers bounded by
$U$, the maximal flow can be found in time $O(\min(n^{2/3} m, m^{3/2})
\log(n^2/m) \log U)$ \cite{gr:maxconstflows}.  For unit networks, the
log-factor is not necessary and the fastest algorithm runs in time
$O(\min(n^{2/3} m, m^{3/2}))$ \cite{et:maxunitflows}.  For undirected unit
networks, the fastest known deterministic algorithm runs in time $O(n^{7/6}
m^{2/3})$ and the fastest known probabilistic algorithm runs in time
$O(n^{20/9})$ \cite{kl:maxunitflows}.

Another well studied problem is finding a matching in a graph.  We are given
an undirected graph.  A matching is a set of edges such that every vertex is
connected to at most one other vertex.  The task is to find a matching of
maximal size.  The classical algorithm based on augmenting paths runs in time $O(n^3)$
\cite{edmonds:matching, gabow:matching}.  If the graph is bipartite, then the
maximal matching can be found in faster time $O(n^{5/2})$
\cite{hk:bipartite-matching}.  Finding a bipartite matching can be reduced to
finding a maximal flow in a directed unit network, hence one can apply the same
algorithms and achieve a running time $O(\min(n^{2/3} m, m^{3/2}))$
\cite{et:maxunitflows}.  Recently, Mucha and Sankowski published a new
algorithm \cite{ms:gauss-matching} based on matrix multiplication that finds a
maximal matching in both bipartite and non-bipartite graphs in time
$O(n^\omega)$, where $2 \le \omega \le 2.38$ is the exponent of the best matrix
multiplication algorithm.

In our paper, we analyze the quantum time complexity of these problems.  We
use Grover's search \cite{grover:search, bbht:bounds} to speed up searching
for an edge.  A similar approach has been successfully applied by D\"urr et
al.~\cite{dhhm:graph} to the following graph problems: connectivity, strong
connectivity, minimum spanning tree, and single source shortest paths.  Our
bipartite matching algorithm is polynomially faster than the best classical
algorithm, its non-bipartite version is faster when the number of edges is $m
= O(n^{1.76-\eps})$ for some $\eps > 0$, and the network flows algorithm is
polynomially faster when $m = \Omega(n^{1+\eps})$ and $U$ is small.

There is an $\Omega(n^{3/2})$ quantum adversary
lower bound for the bipartite matching problem \cite{bdfls:graph, zhang:ambainis}.
Since the bipartite matching problem is a special case of the other
problems studied in this paper, this implies
an $\Omega(n^{3/2})$ quantum lower bound for all 
problems in this paper.

\section{Preliminaries}

An excellent book about quantum computing is the textbook by Nielsen and
Chuang \cite{nc:qc-book}.  In this paper, we only use two quantum
sub-routines and otherwise our algorithm are completely classical.  The first
one is a generalization of Grover's search that finds $k$ items in a
search space of size $\ell$ in total time $O(\sqrt{k
\ell})$~\cite{grover:search, bbht:bounds}.  An additional time $O(\sqrt \ell)$
is needed to detect that there are no more solutions.  The second one is
quantum counting that estimates the number of ones in a string of length $n$
within additive constant $\sqrt n$ with high probability in time $O(\sqrt n)$
\cite[Theorem 13]{bhmt:countingj}.

Each of those algorithms may output an incorrect answer with a constant
probability.  Our algorithms may use a polynomial number $n^c$ of quantum
subroutines.  Because of that, we have to repeat each quantum subroutine
$O(\log n)$ times, to make sure that the probability of an incorrect answer is
less than $1/n^{c+1}$.  Then, the probability that all quantum subroutines in
our algorithm output the correct answer is at least $1-1/n$. This increases
the running time of all our algorithms by a $\log n$ factor.  We omit the
log-factors in the proofs, but we state them in the statements of our
theorems.

A very good book about network flows is the classical book by Ahuja, Magnanti,
and Orlin \cite{amo:network-flows}.  It, however, does not contain most of the
newest algorithms that we compare our algorithms to.  We use the following
concepts:
A \emph{layered network} is a network whose vertices are ordered into a number
of layers, and whose edges only go from the $i$-th layer to the $(i+1)$-th
layer.
A \emph{residual network} is a network whose capacities denote the residual
capacity of the edges in the original network.  When an edge has a capacity
$c$ and carries a flow $f$, then its residual capacity is either
$c-f$ or $c+f$ depending on the direction.
An \emph{augmenting path} in a network is a path from the source to the sink
whose residual capacity is bigger than 0.  An augmenting path for the matching
problem is a path that consists of alternated non-edges and edges of the
current matching, and starts and ends in a free vertex.
A \emph{blocking flow} in a layered residual network is a maximal flow with
respect to inclusion.  A blocking flow cannot be increased by one augmenting
path.
A \emph{cut} in a network is a subset of edges such that there is no path from
the source to the sink if we remove these edges.  The size of a cut is the sum
of the capacities of its edges.  Any flow has size smaller or equal to the
size of any cut.

Let us define our computational model.  Let $V$ be a fixed vertex set of size
$n \ge 1$ and let $E \subseteq ({V \atop 2})$ be a set of edges.  $E$ is a
part of the input.  Let $m$ denote the number of edges.  We assume that $m \ge
1$.  We consider the following two black-box models for accessing directed
graphs:

\begin{itemize}
  \item \emph{Adjacency} model: the input is specified by an $n \times n$
    Boolean matrix $A$, where $A[v,w] = 1$ iff $(v,w) \in E$.
  \item \emph{List} model: the input is specified by $n$ arrays $\{ N_v: v \in
    V \}$ of length $d_v \le n$.  Each entry of an array is either a
    number of a neighbor or an hole, and $\{ N_v[i]: i=1, \dots, d_v \} -
    \{ \mbox{hole} \} = \{ w: (v,w) \in E \}$.
\end{itemize}

The structure of the paper is as follows: In \secref{sec:layers}, we present a
quantum algorithm for computing a layered network from a given network.  It is
used as a tool in almost all our algorithms.  In \secref{sec:bipartite}, we
present a simple quantum algorithm for bipartite matching.  In
\secref{sec:non-bipartite}, we show how to quantize the classical algorithm
for non-bipartite matching.  In \secref{sec:flows}, we present a quantum
algorithm for network flows.

\section{Finding a layered subgraph}
\label{sec:layers}

We are given a connected directed black-box graph $G = (V, E)$ and a starting
vertex $a \in V$, and we want to assign layers $\ell: V \to \N$ to its
vertices such that $\ell(a) = 0$ and $\ell(y) = 1 + \min_{x: (x,y) \in E}
\ell(x)$ otherwise.  The following quantum algorithm computes layer numbers
for all vertices:

\begin{enumerate}
  \item Set $\ell(a) = 0$ and $\ell(x) = \infty$ for $x \ne a$.

    Create a one-entry queue $W = \{ a \}$.
  \item While $W \ne \emptyset$,
    \begin{itemize}
      \item take the first vertex $x$ from $W$,
      \item find by Grover's search all its neighbors $y$ with $\ell(y) =
	\infty$, \\ set $\ell(y) := \ell(x)+1$, and append $y$ into $W$,
      \item and remove $x$ from $W$.
    \end{itemize}
\end{enumerate}

\begin{theorem}
  \label{thm:layers}
  The algorithm assigns layers in time $O(n^{3/2} \log n)$ in the adjacency
  model and in time $O(\sqrt{n m} \log n)$ in the list model.
\end{theorem}

\begin{proof}
The algorithm is a quantum implementation of breadth-first search.  The
initialization costs time $O(n)$.  Every vertex is processed at most once.  In
the adjacency model, every vertex contributes by time at most $O(\sqrt n)$, because
finding a vertex from its ancestor costs time at most $O(\sqrt n)$ and
discovering that a vertex has no descendant costs the same.

In the list model, processing a vertex $v$ costs time $O(\sqrt{n_v d_v} + \sqrt{d_v + 1})$,
where $n_v$ is the number of vertices inserted into $W$ when processing $v$.
Let $f \le \min(n, m)$ be the number of found vertices.
Since $\sum_v n_v \le f \le n$ and $\sum_v (d_v + 1)\le m + f = O(m)$,
the total running time is upper-bounded by the Cauchy-Schwarz inequality as
follows:
\[
\sum_v \sqrt{n_v d_v}
  \le \sqrt{\sum_v n_v} \sqrt{\sum_v d_v}
  = O(\sqrt{n m}),
\]
and $\sum_v \sqrt{d_v+1} \le \sqrt f \sqrt{m + f}$ is upper-bounded in the
same way.
\end{proof}

\section{Bipartite matching}
\label{sec:bipartite}

We are given an undirected bipartite black-box graph $G = (V_1, V_2, E)$ and
we want to find a maximum matching among its vertices.  
This can be done classically in time $O(n^{5/2})$ \cite{hk:bipartite-matching} as follows:

\begin{enumerate}
  \item Set $M$ to an empty matching.
  \item \label{step:layered}
    Let $H = (V',E')$ denote the following graph:
    \begin{eqnarray*}
      V' &=& V_1 \cup V_2 \cup \{ a, b \} \\
      E' &=& \{ (a, x): x \in V_1, (x,\_) \not\in M \} \\
	 &\cup& \{ (x, y): x \in V_1, y \in V_2, (x,y) \in E, (x,y) \not\in M \} \\
	 &\cup& \{ (y, x): x \in V_1, y \in V_2, (x,y) \in E, (x,y) \in M \} \\
	 &\cup& \{ (y, b): y \in V_2, (\_,y) \not\in M \}
    \end{eqnarray*}

    Find a maximal (with respect to inclusion) set $S$ of vertex-disjoint
    augmenting paths of minimal length.  This is done as follows: First,
    construct a layered subgraph $H'$ of $H$.  Second, perform a depth-first
    search for a maximal set of vertex-disjoint paths from $a$ to $b$ in $H'$.
    Every such a path is an augmenting path in $M$, and they all have the same
    minimal length.
  \item Augment the matching $M$ by $S$.
  \item If $S \ne \emptyset$, go back to \stepref{step:layered}, otherwise
    output the matching $M$.
\end{enumerate}

The algorithm is correct because (1) a matching is maximal iff there is no
augmenting path, and (2) the minimal length of an augmenting path is increased
by at least one after every iteration.
The construction of $H'$ classically and the depth-first search both cost
$O(n^2)$.  The maximal number of iterations is $O(\sqrt n)$ due to the
following statement:

\begin{lemma}
  {\rm \cite{hk:bipartite-matching}}
  \label{lem:augm}
  If $M_1$ and $M_2$ are two matchings of size $s_1$ and $s_2$ with $s_1 <
  s_2$, then there exist $s_2 - s_1$ vertex-disjoint augmenting paths in
  $M_1$.
\end{lemma}

Let $s$ be the size of the maximal matching $M$ in $G$, and let $s_i$ be the
size of the found matching $M_i$ after the $i$-th iteration.  Let $j$ be the
number of the first iteration with $s_j \ge s - \sqrt n$.  The total number of
iterations is at most $j + \sqrt n$, because the algorithm finds at least one
augmenting path in every iteration.  On the other hand, by
\lemref{lem:augm}, there are $s - s_j \ge \sqrt n$ vertex-disjoint
augmenting paths in $M_j$.  Since all augmenting paths in the $j$-th
iteration are of length at least $j+2$, it must be that $j < \sqrt n$, otherwise the
paths would not be disjoint.  We conclude that the total number of iterations
is at most $2 \sqrt n$.

\begin{theorem}
  Quantumly, a maximal bipartite matching can be found in time $O(n^2 \log n)$
  in the adjacency model and $O(n \sqrt{m + n} \log n)$ in the list model.
\end{theorem}

\begin{proof}
We present a quantum algorithm that finds all augmenting paths in one
iteration in time $O(n^{3/2})$, resp.\ $O(\sqrt{n (m + n)})$, times a
log-factor for Grover's search.  Since the number of
iterations is $O(\sqrt n)$, the upper bound on the running time follows.  Our
algorithm works similarly to the classical one; it also computes the layered
graph $H'$ and then searches in it.

The intermediate graph $H$ is generated on-line from the input black-box graph
$G$ and the current matching $M$, using a constant number of queries as
follows: the sub-graph of $H$ on $V_1 \times V_2$ is the same as $G$ except
that some edges have been removed; here we exploit the fact that the lists
of neighbors can contain holes.  We also add two new vertices $a$ and $b$, add
one list of neighbors of $a$ with holes of total length $n$, and at most one
neighbor $b$ to every vertex from $V_2$.  \thmref{thm:layers} states how long
it takes to compute $H'$ from $H$.  It remains to show how to find the
augmenting paths in the same time.

This is simple once we have computed the layer numbers of all vertices.  We
find a maximal set of vertex-disjoint paths from $a$ to $b$ by a depth-first
search.  A descendant of a vertex is found by Grover's search over all
unmarked vertices with layer number by one bigger.  All vertices are unmarked
in the beginning.  When we find a descendant,
we mark it and continue backtracking.  Either the vertex will become a part of
an augmenting path, or it does not belong to any and hence it needs not be
tried again.  Each vertex is thus visited at most once.

In the adjacency model, every vertex costs time $O(\sqrt n)$ to be found and
time $O(\sqrt n)$ to discover that it does not have any descendant.  In the
list model, a vertex $v$ costs time $O(\sqrt{n_v d_v} + \sqrt{d_v})$, where
$n_v$ is the number of unmarked vertices found from $v$.  The sum over all
vertices is upper-bounded like in the proof of \thmref{thm:layers}.
Note that $\sum_v d_v$ has been increased by at most $2 n$.
\end{proof}

\section{Non-bipartite matching}
\label{sec:non-bipartite}

We are given an undirected graph $G = (V, E)$ and we want to find a maximal
matching among its vertices.  There is a classical
algorithm~\cite{edmonds:matching, gabow:matching} running in $n$
iterations of time $O(n^2)$.

Each iteration consists of searching for an augmenting path.  The algorithm
performs a breadth-first search from some free vertex.  It browses paths that
consist of alternated non-edges and edges of the current matching.  The
matching is specified by pointers \emph{mate}.  Let us call a vertex $v$
\emph{even} if we have found such an alternated path of even length from the
start to $v$; otherwise we call it \emph{odd}.  Newly discovered vertices are
considered to be odd.  For each even vertex, we store two pointers \emph{link}
and \emph{bridge} used for tracing the path back, and a pointer \emph{first}
to the last odd vertex on this path.  The algorithm works as follows:

\begin{enumerate}
  \item Initialize a queue of even vertices $W = \{ a \}$ with some free
    vertex $a$.
  \item \label{step:takev}
    Take the first vertex $v$ from $W$ and delete it from $W$.
  \item If there exists an free vertex $w$ connected to $v$, then augment the
    current matching by the path $a \to v$ plus the edge $(v,w)$, and quit.  A
    general subpath $\rho: b \to v$ is traced recursively using $v$'s pointers
    as follows:
    \begin{itemize}
      \item If \emph{bridge} is nil, then \emph{link} points to the previous
	even vertex on $\rho$.  Output 2 edges from $v$ to \emph{mate} and \emph{link},
	and trace $\rho$ from \emph{link} to $b$.
      \item Otherwise $v$ was discovered via a bridge, \emph{link} points to
	$v$'s side of the bridge, and \emph{bridge} to the other side.  Trace
	$\rho$ from \emph{link} to $v$ in the opposite direction, and then
	from \emph{bridge} to $b$ in the normal direction.
    \end{itemize}
  \item For every odd vertex $w$ connected to $v$, do the following:
    \begin{itemize}
      \item Let $w$ be connected to a mate $w'$.  If $w'$ is even, do nothing.
      \item Otherwise mark $w'$ as even, append it to $W$, and set its
	pointers as follows: \emph{link} to $v$, \emph{bridge} to nil, and
	\emph{first} to $w$.
    \end{itemize}
  \item For every even vertex $w$ connected to $v$, do the following:
    \begin{itemize}
      \item Compare the pointers \emph{first} of $v$ and $w$.  If they are
	equal, do nothing.
      \item Now, $v$ and $w$ lie on a circle of odd length, and the edge
	$(v,w)$ is a \emph{bridge} between the two subpaths.  Find the nearest
	common odd ancestor $p$ of $v$ and $w$ by tracing the pointers
	\emph{first}.  Collapse the circle as follows:

	\begin{itemize}
	  \item Mark all odd vertices between $v$ and $p$ as even, append them
	    to $W$, and set their pointers as follows: \emph{link} to $v$,
	    \emph{bridge} to $w$, and \emph{first} to $p$.
	  \item Do the same for odd vertices between $w$ and $p$.
	  \item Finally, rewrite all links \emph{first} pointing to odd
	    vertices that have just become even to $p$.
	\end{itemize}
	
    \end{itemize}
  \item If $W$ is empty, then there is no augmenting path from $a$ and we
    quit, otherwise go back to \stepref{step:takev}.
\end{enumerate}

It holds that if an augmenting path from some vertex has not been found, then
it would not be found even later after more iterations of the algorithms.
Hence it suffices to search for an augmenting path from each vertex once.

\begin{theorem}
Quantumly, a maximal non-bipartite matching can be found in time $O(n^{5/2}
\log n)$ in the adjacency model and $O(n^2 (\sqrt{m / n} + \log n) \log n)$ in
the list model.
\end{theorem}

\begin{proof}
The algorithm iteratively augments the current matching by single augmenting
paths, like the classical algorithm.  An augmenting path is found using
Grover's search in faster time $O(n^{3/2})$, resp.\ $O(n (\sqrt{m / n} + \log
n))$, times the usual log-factor.  This implies the bound on the total running
time, since there are $n$ vertices and each of them is used as the starting
vertex $a$ at most once.  Let us prove the time bound for the list model.

Let $f \le \min(n, m)$ denote the number of even vertices.  For every even
vertex $v$, we perform the following 3 Grover's searches:  First, we look for
a free neighbor of $v$ in time $O(\sqrt{d_v})$.  Second, we process all odd
neighbors of $v$ whose mate is still odd in total time $O(\sqrt{e_v d_v})$,
where $e_v$ is the number of odd vertices that are found during processing
$v$.  Third, we process all even neighbors of $v$ whose pointer \emph{first}
is different from $v$'s pointer \emph{first}, in time $O(\sqrt{b_v d_v})$,
where $b_v$ is the number of bridges that are found during processing $v$.
Clearly $\sum_v e_v \le f$ and $\sum_v b_v \le f$, and, since $\sum_v d_v \le
m$, by the Cauchy-Schwarz inequality, the total time spent in all Grover's
searches is $O(\sqrt{n m})$.

Let us estimate the running time of collapsing one circle.  Let $p_1$ be the
length of the link-list of pointers \emph{first} from one side of the bridge
into the nearest common parent, let $p_2$ be the other one, and let $p =
\max(p_1, p_2)$.  The nearest common parent is found in time $O(p \log p)$ as
follows: we maintain two balanced binary trees for each link-list, add
vertices synchronously one-by-one, and search for every newly inserted vertex
in the opposite tree, until we find a collision.  Let $r_v$ be the number of
odd vertices collapsed during processing a vertex $v$.  It holds that $r_v =
p_1 + p_2 = \Theta(p)$ and $\sum_v r_v \le f$.  Hence the total time spent in
collapsing circles is $O(f \log f)$.

Rewriting the pointers \emph{first} of all even vertices inside a collapsed
circle would be too slow.  We instead maintain aside a Union-tree of all these
pointers, and for every odd vertex converted to even, we append its subtree to
the node of the nearest common ancestor.  The total time spent in doing this
is $O(f \log f)$.

The augmenting path has length at most $n$ and it is traced back in linear
time.  We conclude that the total running time of finding an augmented time is
$O(\sqrt{n m} + n \log n) = O(n (\sqrt{m / n} + \log n))$, which is $O(\sqrt{n
m})$ for $m \ge n (\log n)^2$.  The running time in the adjacency model is equal
to the running time in the list model with $m = n^2$, that is $O(n^{5/2})$.
\end{proof}

\section{Integer network flows}
\label{sec:flows}

We are given a directed network with real capacities, and we want to find
a maximal flow from the source to the sink.  There are classical algorithms
running in time $O(n^3)$ \cite{karzanov:preflows, 3inds:maxflows}.  They
iteratively augment the current flow by adding blocking flows in
layered residual networks~\cite{dinic:layers} of increasing depth.
Since the depth is
increased by at least one after each iteration, there are at most $n$
iterations.  Each of them can be processed in time $O(n^2)$.  For sparse
real networks, the fastest known algorithm runs in time $O(n m (\log n)^2)$
\cite{gn:maxflows}.

Let us restrict the setting to integer capacities bounded by $U$.  There is a
simple capacity scaling algorithm running $O(n m \log U)$ \cite{ek:maxflows,
dinic:layers}.  The fastest known algorithm runs in time $O(\min(n^{2/3} m,
m^{3/2}) \log(n^2/m) \log U)$ \cite{gr:maxconstflows}.
For unit networks, i.e.\ for $U=1$, a simple combination of the capacity
scaling algorithm and the blocking-flow algorithm runs in time $O(\min(n^{2/3}
m, m^{3/2}))$ \cite{et:maxunitflows}.  For undirected unit networks, there is
an algorithm running in time $O(n^{3/2} \sqrt m)$ \cite{gr:maxunitflows}, and
the fastest known algorithm runs in worst-case time $O(n^{7/6} m^{2/3})$
and expected time $O(n^{20/9})$ \cite{kl:maxunitflows}.

\begin{lemma}
  {\rm \cite{et:maxunitflows}}
  \label{lem:residual}
  Let us have an integer network with capacities bounded by $U$, whose layered
  residual network has depth $k$.  Then the size of the residual flow is at
  most $\min( (2n/k)^2, m/k ) \cdot U$.
\end{lemma}

\begin{proof}
\begin{enumerate}
  \item[(1)]
    There exist layers $V_\ell$ and $V_{\ell+1}$ that both have less than
    $2n/k$ vertices.  This is because if for every $i=0, 1, \dots, k/2$, at
    least one of the layers $V_{2i}, V_{2i+1}$ had size at least $2n/k$,
    then the total number of vertices would exceed $n$.  Since $V_\ell$ and
    $V_{\ell+1}$ form a cut, the residual flow has size at most $|V_\ell|
    \cdot |V_{\ell+1}| \cdot U \le (2n/k)^2 U$.

  \item[(2)]
    For every $i=0, 1, \dots, k-1$, the layers $V_i$ and $V_{i+1}$ form a cut.  
    These cuts are disjoint and they together have at most $m$ edges.  Hence
    at least one of them has at most $m/k$ edges, and
    the residual flow has thus size at most $O(m U / k)$.
\end{enumerate}
\end{proof}

\begin{theorem}
  \label{thm:maxflow}
Let $U \le n^{1/4}$.
Quantumly, a maximal network flow with integer capacities at most $U$ can be
found in time $O(n^{13/6} \cdot U^{1/3} \log n)$ in the adjacency model and in time
$O(\min(n^{7/6} \sqrt m \cdot U^{1/3}, \sqrt{n U} m) \log n)$ in the list model.
\end{theorem}

\begin{proof}
The algorithm iteratively augments the current flow by blocking flows in
layered residual networks~\cite{dinic:layers}, until the depth of the network
exceeds $k = \min(n^{2/3} U^{1/3}, \sqrt{m U})$.  Then it switches to searching single
augmenting paths~\cite{ek:maxflows}, while there are some.  The idea of
switching the two algorithms comes from \cite{et:maxunitflows}.  Our algorithm
uses classical memory of size $O(n^2)$ to store the current flow and its
direction for every edge of the network, and a 1-bit status of each vertex.  A
blocking flow is found as follows:

\begin{enumerate}
  \item Compute a layered subgraph $H'$ of the \emph{residual} network
    $H$.  The capacity of each edge in $H$ is equal to the
    original capacity plus or minus the current flow depending on the
    direction.  Edges with zero capacities are omitted.
  \item Mark all vertices as enabled.
  \item \label{step:augm}
    Find by a depth-first search a path $\rho$ in $H'$ from the source to the
    sink that only goes through enabled vertices.  If there is no such a path,
    quit.  During back-tracking, disable all vertices from which there is no
    path to the sink.
  \item Compute the minimal capacity $\mu$ of an edge on $\rho$. \\
    Augment the flow by $\mu$ along $\rho$.
  \item Go back to \stepref{step:augm}.
\end{enumerate}

The layered subgraph $H'$ is computed from $H$ using \thmref{thm:layers}, and
the capacities of $H$ are computed on-line in constant time.  When the flow is
augmented by $\mu$ along the path $\rho$, the saturated edges will have been
automatically deleted.  This is because the algorithm only stores layer
numbers for the vertices, and the edges of $H'$ are searched on-line by
Grover's search.

Let us compute how much time the algorithm spends in a vertex $v$ during
searching the augmenting paths.  Let $a_v$ denote the number of augmenting
paths going through $v$ and let $e_{v,i}$ denote the number of outgoing
edges from $v$ at the moment when there are still $i$ remaining augmenting
paths.  The capacity of every edge is at most $U$, hence $e_{v,i} \ge \lceil i
/ U \rceil$.  The time spent in Grover's searches leading to an augmenting
path in $v$ is thus at most
\[
\sum_{i=1}^{a_v} \sqrt{d_v \over e_{v,i}}
\le \sqrt U \cdot \sum_{i=1}^{a_v} \sqrt{d_v \over i}
= O(\sqrt{U a_v d_v}).
\]
Let $c_v$ denote the number of enabled vertices found from $v$ that do not lie
on an augmenting path and are thus disabled.  The time spent in Grover's
searches for these vertices is at most $O(\sqrt{c_v d_v})$.  Furthermore,
it takes additional time $O(\sqrt{d_v + 1})$ to discover that there is no
augmenting path from $v$, and in this case $v$ is disabled and never visited
again.

Let $j$ denote the depth of the network and let $A_j$ be the size of its
blocking flow.  The total number of augmenting paths going through vertices in
any given layer is at most $A_j$.  We conclude that $\sum_v a_v \le j A_j$.
We also know that $\sum_v c_v \le n$.
Since $\sum_v d_v \le m$, by the Cauchy-Schwarz inequality, the total time
spent by finding one blocking flow is
\begin{eqnarray*}
\sum_v (\sqrt{U a_v d_v} + \sqrt{c_v d_v} + \sqrt{d_v+1})
&\le& \sqrt U \sqrt{\sum_v a_v} \sqrt{\sum_v d_v} + 2 \sqrt{n m} \\
&=& O(\sqrt{j m A_j U} + \sqrt{n m}).
\end{eqnarray*}

Our algorithm performs at most $k = \min(n^{2/3} U^{1/3}, \sqrt{m U})$ iterations of
finding the blocking flow in total time at most $\sqrt{m U} \cdot \sum_{j=1}^k
\sqrt{j A_j} + k \sqrt{n m}$.  Let us assume that the algorithm has not
finished, and estimate the size of the residual flow and thus upper-bound the
number of augmenting paths that need to be found.  The algorithm has
constructed in this iteration a layered network of depth bigger than $k$.  By
\lemref{lem:residual}, the residual flow has size $O(\min( (n/k)^2, m/k )
\cdot U) = O(k)$,
hence the algorithm terminates in $O(k)$ more iterations.  From
this point on, the algorithm only looks for one augmenting path in each
layered network, hence its complexity drops to $O(\sqrt{j' m}) = O(\sqrt{n m})$ per iteration,
omitting the factor $\sqrt{A_{j'} U}$.  The total running time is thus at most
\[
O \Big(
\sqrt{m U} \cdot \sum_{j=1}^k \sqrt{j A_j}
  + k \sqrt{n m}
  \Big)
  + O(k \sqrt{n m}).
\]
Let us prove that $\sum_j \sqrt{j A_j} = O(k^{3/2})$.  We split the sequence
into $\log k$ intervals $S_i = \{ 2^i, 2^i + 1, \dots, 2^{i+1}-1 \}$ of length
$2^i$.  By \lemref{lem:residual}, the residual flow after $\ell = k/2^i$
iterations is at most $O(\min( (n/k)^2 \cdot 2^{2 i}, m/k \cdot 2^i) \cdot U)
\le O(2^{2 i} k) = O((k/\ell)^2 \ell) = O(k^2 / \ell)$.  Since the total size
of all blocking flows cannot exceed the residual flow, $\sum_{j=\ell}^{2 \ell
- 1} A_j = O(k^2 / \ell)$.  By applying the Cauchy-Schwarz inequality
independently on each block, we get
\begin{eqnarray*}
\sum_{j=1}^k \sqrt{j A_j}
&=& \sum_{i=0}^{\log k} \sum_{j=2^i}^{2^{i+1}-1} \sqrt{j A_j}
\le \sum_{i=0}^{\log k} \sqrt{2^i \cdot 2^{i+1}} \sqrt{ \sum_{j=2^i}^{2^{i+1}-1} A_j}
\\
&\le& \sqrt 2 \sum_{i=0}^{\log k} 2^i \sqrt{k^2 / 2^i}
= \sqrt 2 \cdot k \sum_{k=0}^{\log k} 2^{i/2}
= O(k^{3/2}).
\end{eqnarray*}
The total running time is thus $O(k \sqrt m (\sqrt{k U} + \sqrt n))$.  Now, $k
U \le n$, because $U \le n^{1/4}$ and $k U = \min( n^{2/3} U^{4/3}, \sqrt m
\cdot U^{3/2}) \le n^{2/3} n^{1/3} = n$.  The running time is therefore $O(k
\sqrt{n m}) = O(\min(n^{7/6} \sqrt m \cdot U^{1/3}, \sqrt{n U} m))$, times a
log-factor for Grover's search.  The time for the adjacency model follows from
setting $m = n^2$ and it is $O(n^{13/6} \cdot U^{1/3} \log n)$.
\end{proof}


It is not hard to compute an upper bound on the running time of the network
flows algorithm for $U > n^{1/4}$ by the same techniques.  One obtains
$O(\min(n^{7/6} \sqrt m, \sqrt n m) \cdot U \log n)$ for arbitrary $U$ by
setting $k = \min(n^{2/3}, \sqrt m)$.  It would be interesting to apply
techniques of~\cite{gr:maxconstflows} to improve the multiplicative constant
in \thmref{thm:maxflow} from $\mathrm{poly}(U)$ to $\log U$.  If $m =
\Omega(n^{1+\eps})$ for some $\eps > 0$ and $U$ is small, then our algorithm
is polynomially faster than the best classical algorithm.  For constant $U$
and $m = O(n)$, it is slower by at most a log-factor.  The speedup is biggest
for dense networks with $m = \Omega(n^2)$.

\begin{theorem}
Any bounded-error quantum algorithm for network flows with integer capacities
bounded by $U = n$ has quantum query complexity $\Omega(n^2)$.
\end{theorem}

\begin{proof}
Consider the following layered graph with $m = \Theta(n^2)$ edges.  The
vertices are ordered into 4 layers: the first layer contains the source, the
second and third layer contain $p = \frac n2 - 1$ vertices each, and the last
layer contains the sink.  The source and the sink are both connected to all
vertices in the neighboring layer by $p$ edges of full capacity $n$.  The vertices
in the second and third layer are connected by either $\frac{p^2}2$ or
$\frac{p^2}2 + 1$ edges of capacity 1 chosen at random.  The edges between
these two layers form a minimal cut.  Now, deciding whether the maximal flow
is $\frac{p^2}2$ or $\frac{p^2}2 + 1$ allows us to compute the majority on
$p^2$ bits.  There is an $\Omega(p^2) = \Omega(n^2)$ lower bound for majority,
hence the same lower bound also holds for the computation of the maximal flow.
\end{proof}

\section*{Acknowledgments}

We thank Marek Karpinski for discussions that lead to
the quantum bipartite matching algorithm and help with literature on
classical algorithms for bipartite matchings.

\bibliographystyle{alpha}
\bibliography{../quantum}

\end{document}